\documentclass[twocolumn,showpacs,preprintnumbers,amsmath,amssymb]{revtex4}

\usepackage{graphicx}
\usepackage{dcolumn}
\usepackage{bm}

\def\be{\begin{equation}}
\def\ee{\end{equation}}
\def\bea{\begin{eqnarray}}
\def\eea{\end{eqnarray}}

\begin{document}

\preprint{xxxx}

\title{Relativistic Quantum State, Geometric Phase and Dissipationless Spin current 

}

\author{S. Arunagiri}\email{arunairi@postech.ac.kr}

\affiliation{Division of Advanced Materials Science, Pohang University of Science and Technology,\\
San 31, Hyoja Dong, Nam Gu, Pohang 790 784, South Korea}

\date{\today}

\begin{abstract}
Spin current of a Dirac particle is shown to be given by the geometric phase and in terms of the later, a closed form expression is obtained for the dissipationlessness of the spin current.
\end{abstract}

\pacs{03.65.Pm, 11.40.-q, 71.70.Ej, 75.76.+j, 85.75.-d}
\keywords{Dirac Equation, Spin current, Spintronics}
\maketitle
The emerging science of spintronics is based on spin current. Like electric charge current, the spin current is also a macroscopic phenomenon in many body systems. However, unlike the charge, the spin is a quantum property with no classical analogue. The dynamics that underlies the spin current is the spin-orbit interaction \cite{dresselhaus, rashbha}. An intriguing aspect of the spin transport is that the spin current is dissipationless: sum of the oppositely polarised spin currents does not vanish. This is an effect of collective and coherent motion of spin in many body dynamics with potential consequences  \cite{murakami}. The spin-orbit coupling gives rise to new phenomena such as giant magnetoresistence and spin Hall effect and other applications \cite{dsarma}. In materials such as graphene and topological insulators, the electrons exhibit relativistic-like character. These same materials are also subject of spintronics pursuit. The dynamics of spin-orbit interaction and the dissipationless nature of spin current in many body systems do present in the single particle description, albeit its small magnitude. Then it is interesting to understand the spin current and its properties from a fundamental point of view \cite{sun, vernes, dartora, sobreiro}. 

A definition of spin current, in the Dirac description of  electron, is given by the Gordon decomposition which states that the Dirac current is sum of the spatial and spin currents \cite{bjorken}. While in the case of the spatial current, the initial and final states are the same, either upper or lower components of Dirac spinor, the spin current involves different components of the spinor as intial and final states. Otherwise, the spin current would vanish. The decomposition is valid for massive electron. This definition of spin current reflects spin-orbit coupling as well. 

The purpose of this letter is to investigate further into the definition of spin current and its disspationless property in the Dirac theory. We show that the spin current is related to adiabatic evolution of the spin state giving rise to geometric phase. Further we obtain a closed form expression for the dissipationless of spin current in terms of the geometric phase. 

Let us consider the commutation relation between the Dirac Hamiltonian, $H = c\gamma^\mu p_\mu + mc^2$ and $\gamma^\nu$  
(in the standard notation \cite{bjorken})
\be
\left[ H, \gamma^\nu \right] = - 2 i \sigma^{\mu\nu} p_\mu \label{hgamcomm}
\ee
This relation is independent of electron mass. We could take this commutation relation writing the Hamiltonian $H$ and the current $\gamma^\nu$ in terms of fermionic fields as well. The result would be the same in form and content. Subsequently, the equation of motion for $\gamma^\nu$ is
\be
\dot{\gamma}^\nu  = {2 \over \hbar}c \sigma^{\mu\nu} p_\mu \label{eomgamma}
\ee
with $ \dot{\gamma}^\nu $ refers to the time derivative.
Multiplying both sides of the eq. (\ref{eomgamma}) by $\vert \psi_i \rangle$ from the right and by $\langle \psi_j \vert $ from the left, we get
\be
\langle \psi_j \vert \dot{\gamma}^\nu\vert \psi_i \rangle  = 
{2 \over \hbar}c \langle \psi_j \vert \sigma^{\mu\nu} p_\mu  \vert \psi_i \rangle \label{eomgamma1}
\ee
where $\vert \psi_i \rangle$ and $\vert \psi_j \rangle$ are solutions to the Dirac equation and the subscripts $i, j$ refers to initial and final state.  From (\ref{eomgamma}) and (\ref{eomgamma1}), denoting $j^\nu_s =  \langle \psi_j \vert \sigma^{\mu\nu} p_\mu  \vert \psi_i \rangle$ with the subscript $s$ referring to spin polarisations $\pm$, we obtain the spin current as 
\be
j^\nu_s = - {\hbar \over {c}} \langle \dot{\psi_j} \vert \gamma^\nu \vert \psi_i \rangle \label{js1}
\ee
The eq. (\ref{js1}) is an alternate definition for spin current. Note the time derivative of $\vert \psi \rangle$. In the above, we used: 
$\langle \psi_j \vert \dot{\gamma}^\nu \vert \psi_i \rangle = 
d \left( \langle \psi_j \vert \gamma^\nu \vert \psi_i \rangle \right)/dt
- \langle \dot{\psi_j} \vert \gamma^\nu \vert \psi_i \rangle
- \langle \psi_j \vert \gamma^\nu \vert \dot{\psi_i} \rangle$ and the later two terms on the rhs are equal to one another and the first one vanishes. 

The RHS of the equation above (\ref{js1}) is determined considering the evolution of $\vert \psi_j \rangle$ adiabatically. In non-relativistic quantum mechanics, the adiabatic evolution is associated with geometric phase \cite{berry} with holonomy properties \cite{simon}. It is shown that geometric phase can also be associated with both Klein-Gordon \cite{anandan} and Dirac fields \cite{wang}. In respect of the relativistic fermion, the geometric phase is shown to be covariant and each component of the spinor is associated with its own phase. This implies that the adiabatic evolution of Dirac spinor does not result in a final state admixture positive and negative energy components. 

\begin{figure}[t]
\centering%
\includegraphics[scale=0.5]{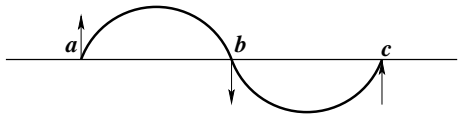}
\caption{Spin current as Berry curvature: $A_0$ vs. period (time)}
\label{fig:figure1}
\end{figure}

The evolution of $\vert \psi_i \rangle$ that is related to the geometric phase is given by, as in the case of Schrodinger Hamiltonian,:
\be
\vert \dot{\psi}_i \rangle = U \vert \psi_i \rangle \label{gphase}
\ee
where 
\be
U = \exp\left[ -i \int A_\mu x^\mu \right] \label{phase}
\ee
The exponent $\int A_\mu x^\mu = \phi$ the geometric phase associated with adiabatic evolution of $\vert \psi_i \rangle$
with the emergence of the geometric phase as gauge field, $A_\mu$ \cite{wilczek}. Using (\ref{gphase}) and (\ref{phase}), the spin current expression in (\ref{js1}) becomes as
\be
j^\nu_+ = - {\hbar \over {c}} \left(\gamma^\nu\right)_{ji} \langle \psi_j \vert U^\dagger \vert \psi_i \rangle \label{jsp}
\ee
where $+$ denotes the polarisation. Similarly, reversing the spin direction, the corresponding spin current is
\be
j^\nu_- = - {\hbar \over {c}} \left(\gamma^\nu\right)_{ji} \langle \psi_j \vert U \vert \psi_i \rangle \label{jsm}
\ee
The sum of the oppositely polarised currents, (\ref{jsp}) and (\ref{jsm}), is then
\be
j^\nu_+ + j^\nu_- = - {\hbar \over {c}} \left(\gamma^\nu\right)_{ji} \langle \psi_j \vert \psi_i \rangle \left(U^\dagger + U \right) \label{jsd1}
\ee
The sum that does not vanish is known as dissipationless property of the spin current: 
\be
j^\nu_+ + j^\nu_- = - 2 {\hbar \over {c}} \left(\gamma^\nu\right)_{ji} \langle \psi_j \vert \psi_i \rangle \cos \phi \label{jsd1}
\ee
This is our closed form expression for the spin current dissipationlessness which is proportional to cosine of the geometric phase. It is not vanishing in the limit of $\phi = 0$. The phase $\phi$ is function of time. Adiabatic evolution of the fermion field along a closed loop yields the curvature shown in the figure \ref{fig:figure1}. 

In conclusion, it is showed that the spin current obtained out of the commutation proprties of the dirac Hamiltonian and $\gamma^\nu$ leads to an alternate definition of spin current in terms of geometric phase and the dissipationelessness of the spin current is shown to be proportional to the cosine of the geometric phase. 

\acknowledgements
Part of this work was done at the Institute of Mathematical Sciences, Chennai. The author is grateful to G. Baskaran, H. S. Sharatchandra and S. Kettemann for useful discussions. This research is supported by WCU(World Class University) program through the National Research Foundation of Korea funded by the Ministry of Education, Science and Technology(R31-2008-000-10059-0).

\end{document}